\documentclass[twocolumn,nofootinbib,showpacs,amsmath,amssymb]{revtex4}
\usepackage{graphicx}
\def\dd{\displaystyle}
\begin{document}
\title{\bf A new renormalization approach to the Dirichlet Casimir
effect for $\phi^4$ theory \\ in (1+1) dimensions}
\author{Reza Moazzemi}
\email{email: R-Moazzemi@sbu.ac.ir}
\author{Siamak S. Gousheh}%
\affiliation{%
Department of Physics, Shahid Beheshti University, Evin, Tehran
19839, Iran
}%
\date{\today}

\begin{abstract}
The next to the leading order Casimir effect for a real scalar
field, within $\phi^4$ theory, confined between two parallel plates
is calculated in one spatial dimension. Here we use the Green's
function with the Dirichlet boundary condition on both walls. In
this paper we introduce a systematic perturbation expansion in which
the counterterms automatically turn out to be consistent with the
boundary conditions. This will inevitably lead to nontrivial
position dependence for physical quantities, as a manifestation of
the breaking of the translational invariance. This is in contrast to
the usual usage of the counterterms, in problems with nontrivial
boundary conditions, which are either completely derived from the
free cases or at most supplemented with the addition of counterterms
only at the boundaries. We obtain \emph{finite} results for the
massive and massless cases, in sharp contrast to some of the other
reported results. Secondly, and probably less importantly, we use a
supplementary renormalization procedure in addition to the usual
regularization and renormalization programs, which makes the usage
of any analytic continuation techniques unnecessary.
\end{abstract}

\maketitle

\section{Introduction}
\baselineskip=.45cm
 During the last fifty years many papers have
been written on the calculation of the Casimir energy. In this paper
we introduce a new approach in regards to the renormalization
program. We found it most suitable to introduce our approach in the
simplest example possible, i.e. a real scalar field confined between
two parallel plates in 1+1 dimensions, with $\phi^4$
self-interaction. As we shall see, our results for the next to
leading order term (NLO) are finite for both the massive and
massless cases and this differs significantly from what exists in
the literature. It is therefore suitable to start at the beginning.
In 1948 H.B.G. Casimir found a simple yet profound explanation for
the retarded van der Waals interaction \cite{Casimir}. After a short
time, he and D. Polder related this effect to the change in the zero
point energy of the quantum fields due to the presence of nontrivial
boundary conditions \cite{Polder}. This energy has since been called
the Casimir energy. The zero-order energy in perturbation theory has
been calculated for various fields (see for example \cite{Svaiter}).
Also the NLO correction, which is usually called the first-order
effect, has been computed for various fields. For the
electromagnetic field this correction is said to be due to the
following Feynmann diagram \raisebox{-2mm}{\includegraphics{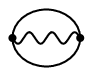}} ,
and has been computed first by Bordag and collaborators
\cite{brw,bl,rsw,bs,sx,zx}. However, note that this correction is a
two loop correction in this case and is ${\cal O}(e^2)$. Moreover
the two-loop radiative corrections for some effective field theories
have been investigated in \cite{rt,kr,km}. Next, in the case of a
real massive scalar field NLO correction to the energy has been
computed in
\cite{ford,kay,toms,lang,albu,Mostepanenko,Baron1,Baron2}. This
correction is a two loop correction in this case but is $\cal
O(\lambda)$. Moreover, N. Graham \emph{et al} used new approaches to
this problem by utilizing the phase shift of the scattering states
\cite{Graham}, or replacing the boundary conditions by an
appropriate potential term  \cite{Graham2}. However, the authors use
the free counterterms, by which we mean the ones that are relevant
to the free cases with no nontrivial boundary conditions, and are
obviously position independent. Only in Ref. \cite{albu} the author notes
that in certain cases, counterterms can depend on the distance
between the plates. The first use of nontrivial boundary conditions
for the renormalization programs in problems of this sort seems to
be due to Fosco and Svaiter \cite{Fosco}. These authors use free
counterterms in the space between the plates and place additional
surface counterterms at the boundaries. Later on various authors
proposed the use of exactly the same renormalization procedure for
various physical problems \cite{Nogu}. The first calculation for the
NLO of Casimir energy for the massive scalar field using this
renormalization program is done in Ref. \cite{Caval}. We should note
that their results for the massless limit in 1+1 dimensions, like
those of \cite{Baron1}, is infinite. It is also worth mentioning
that all the papers on the analogous calculations of the NLO
corrections to the mass of solitons, that we are aware, of use free
counterterms (see for example \cite{soliton,rebhan,dashen}). In
references \cite{rebhan} the authors used the mode number cutoff
introduced by R.F. Dashen (1974) \cite{dashen} to calculate the NLO
Casimir energy due to the presence of solitons.

In this paper, we present a systematic approach to the
renormalization program for problems which are amenable to
renormalized perturbation theory, and contain either nontrivial
boundary conditions or nontrivial (position dependent) backgrounds,
e.g. solitons, or both. Obviously all the n-point functions of the
theory will have in general nontrivial position dependence in the
coordinate representation. This is one of the manifestations of the
breaking of the translational symmetry. The procedure to deduce the
counterterms from the n-point functions in a renormalized
perturbation theory is standard and has been available for over half
a century. Using this, as we shall show, we will inevitably obtain
position dependent counterterms. Therefore, the radiative
corrections to all the input parameters of the theory, including the
mass, will be in general position dependent. Therefore, we believe
the information about the nontrivial boundary conditions or position
dependent backgrounds are carried by the full set of n-point
functions, the resulting counterterms, and the renormalized
parameters of the theory. Our preliminary investigations have
revealed that our position dependent counterterms approach the free
ones when the distance between the plates is large. Their main
difference is for positions which are about a Compton wavelength
away from the walls, although it is also nontrivial at other places.
Here we use this procedure to compute the first-order radiation
correction to the Casimir energy for a real scalar field in 1+1
dimensions. We compute this correction for both a massive and a
massless scalar fields and show that the massless limit of the
massive case exactly corresponds to the massless case.

In addition, up to now all the papers on the Casimir effect, that
we are aware of, use some from of \emph{analytic continuation}. We
share the point of view with some authors such as the ones in
\cite{kay,Mostepanenko} that the analytic continuation techniques
are not always completely justified physically. Moreover, like the
first of the aforementioned authors, we have found counterexamples,
which we point out in this paper and elsewhere \cite{moazzemi2}. The
counterexamples show that it alone might not yield correct physical
results, and sometimes even gives infinite results \cite{Bender}.
Therefore, we prefer to use a completely physical approach by
enclosing the whole system in a box of linear size $L$, which
eventually can go to infinity, and calculating the difference
between the zero point energies of two different configurations. The
main idea of this method is actually due to T.H. Boyer
\cite{Boyer}, who used spheres instead of boxes. This we shall call
the ``box renormalization scheme" and can be used as a supplementary
part of other usual regularization or renormalization programs. This
box renormalization scheme, has the following advantages:
\begin{enumerate}
    \item Use of this procedure removes all of the ambiguities associated
    with the appearance of the infinities, and we use the usual prescription for
    removing the infinities in the regulated theory, as explained in Sec. \ref{massive case}. This is all done without
    resorting to any analytic continuation schemes.
    \item The infrared divergences which generically appear in
    these problems in 1+1 dimensions automatically cancel each
    other.
    \item In order to calculate the Casimir energy we subtract
    two physical configuration of similar nature, e.g. both
    confined within finite regions, and not one confined and the other in
    an unbounded region.
    \item This method can be used as a check for the cases where
    analytic continuation yields finite results, and more
    importantly, can be used to obtain finite results when the
    former yields infinite results.
\end{enumerate}
we should mention that some authors believe that use of box
regularization or renormalization procedures, in which the size of
the box eventually goes to infinity could be avoided by using
appropriate boundary conditions on the fields at spatial infinity
\cite{Nest}

In Section \ref{sec2} we calculate the leading order term for the
Casimir energy in $d$ space dimensional case. We do this first of
all to explain more completely the physical content of the problem
and set up our notations. Secondly this computation is just about as
easy as to do in $d$ dimensions  as  is in the one dimensional case.
In Section \ref{sec3} we compute the first order radiative
correction to this energy. In order to do this we first state the
renormalization conditions, and then derive expressions for the the
first order radiative corrections for both the massive and massless
cases. We show that the results for the massless limit of the
massive case and the massless case are equal. In Section \ref{sec4}
we summarize our results and state our conclusions.

\section{The Leading Term of the Casimir Effect}\label{sec2}
The lagrangian density for a real scalar field with $\phi^4$
self-interaction is:
\begin{equation}\label{e1}
  {\cal L}(x)
  =\frac{1}{2}[\partial_{\mu}\varphi(x)]^{2}-\frac{1}{2}m_0^{2}\varphi(x)^{2}
  -\frac{\lambda_{0}}{4!}\varphi(x)^{4},
\end{equation}
where $m_{0}$ and $\lambda_{0}$ are the bare mass and bare coupling
constant, respectively. Here we calculate the leading term for the
Casimir energy in $d$ spatial dimensions. Obviously the leading
term, in contrast to the higher order corrections, is independent of
the form of the self-interaction. The Casimir energy is in general
equivalent to the work done on the system for bringing two parallel
plates from $\pm \infty$ to $\pm a/2$. As mentioned before, part of
our renormalization procedure is to enclos the whole system in a
$d$ dimensional cubical box of sides $L$. To compute this leading
term, we first compare the energies in two different configurations:
when the plates are at $\pm a/2$ as compare to $\pm b/2$. We name
the axis perpendicular to the plates the $z$ axis. To keep the
expressions symmetrical, we choose the coordinates so that the edges
of the confining box are at $\pm L/2$ in any direction.

\begin{figure}[th]
\begin{center} \includegraphics[width=7cm]{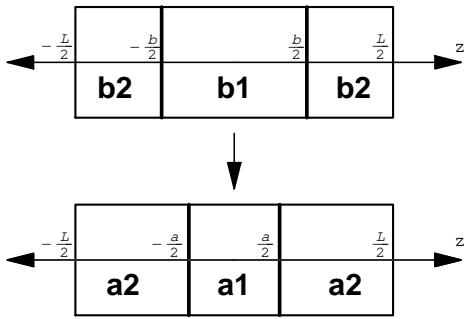}\caption{\label{fig:1} \small
The geometry of the two different configurations whose energies are
to be compared. The labels a1, {\em etc.} denote the appropriate
sections in each configuration separated by the plates.}
\label{geometry}
\end{center}
\end{figure}

The total zero point energy of the upper configuration in figure
(\ref{geometry}) will be called $E_{b}$ and of the lower one
$E_{a}$. In our box renormalization scheme we need to define the
Casimir energy as follows

\begin{equation}\label{e2:E.Cas.}
   E_{\mbox{\tiny \mbox{\tiny Cas.}}}=\lim_{b/a\rightarrow\infty}\left[\lim_{L/b\rightarrow\infty}
   \left(E_{a}-E_{b}\right)\right],
\end{equation}
where,
\begin{equation}\label{e3}
   E_{a}=E_{a_1}+2E_{a_2},\quad
   E_{b}=E_{b_1}+2E_{b_2}.
\end{equation}
Here we choose the Dirichlet boundary condition on the plates. Then
we can expand the field operator $\varphi$ in the eigenstate basis
appropriate to this boundary condition, and its explicit second
quantized form, for example in region $a_1$ becomes
\begin{eqnarray}\label{e4}
   \varphi_{_{a_1}}(x)&=&\int\frac{d^{d-1}\mathbf{k}^{\bot}}{(2\pi)^{d-1}}
   \sum_{n=1}^{\infty}\left(\frac{1}{a\omega_{a_1,n}}\right)^{1/2}\nonumber\\
   &\times&\Bigg\{e^{-i(\omega_{a_1,n}t-\mathbf{k^{\bot}}\textbf{.}
   \mathbf{x}^{\bot})}\sin\left[k_{a_1,n}(z+\frac{a}{2})\right]\textbf{a}_{n}+\nonumber\\
   && \dd e^{i(\omega_{a_1,n}t-\mathbf{k^{\bot}}\textbf{.}
   \mathbf{x}^{\bot})}\sin\left[k_{a_1,n}(z+\frac{a}{2})\right]
   \textbf{a}_{n}^{\dag}\Bigg\},
\end{eqnarray}
where,
\begin{eqnarray}\label{e5}
   \omega_{a_1,n}^{2}&=&m_0^{2}+{k^{\bot}}^{2}+k_{a_1,n}^{2},\nonumber\\
   k_{a_1,n}&=&\frac{n
   \pi}{a} \quad \mbox{and}\quad n=1,2,\ldots.
\end{eqnarray}
Here $\mathbf{k}^{\bot}$ and $k_{a_1,n}$ denote the momenta parallel
and perpendicular to plates (in z-direction), respectively. Also
$\textbf{a}_{n}^{\dag}$ and $\textbf{a}_{n}$ are creation and annihilation
operators  obeying the usual commutation relations:
\[[\textbf{a}_{n},\textbf{a}^{\dag}
_{n^{_\prime}}]=\delta_{n,n^{_\prime}},\quad
[\textbf{a}_{n},\textbf{a}_{n^{_\prime}}]=[\textbf{a}^{\dag}_{n},
\textbf{a}^{\dag}_{n^{_\prime}}]=0,\] and $\textbf{a}|0\rangle=0$
defines the vacuum state in the presence of boundary conditions.
Using the above equations one can easily obtain
\begin{eqnarray}\label{e6}
   E^{(0)}_{a_1}&=&\dd\int d^{d}\textbf{x}\dd\langle0|{\mathcal H}^{(0)}|0\rangle
    = L^{d-1}\int\frac{d^{d-1}
   \mathbf{k}^{\bot}}{(2\pi)^{d-1}}\sum_{n=1}^{\infty}\frac{\omega_{a_1,n}}{2}\nonumber\\
   & =&\frac{L^{d-1}}{2}\frac{\Omega_{d-1}}{(2\pi)^{d-1}}
   \int^{\infty}_{0}dk k^{d-2}\sum_{n=1}^{\infty}\omega_{a_1,n},
\end{eqnarray}
where ${\mathcal H}^{(0)}$ denotes the usual free Hamiltonian
density, easily obtained from the Lagrangian density, and the
superscript $(0)$ denotes the zero (or leading) order term of this
energy. Also $k=|\mathbf{k}^{\bot}|$, and
$\Omega_d=\dd\frac{2\pi^{d/2}}{\Gamma(\frac{d}{2})}$ is the solid
angle in d-dimensions. Therefore,
\begin{equation}\label{e7}
   E^{(0)}_{a}-E^{(0)}_{b}=\frac{L^{d-1}}{2}\frac{\Omega_{d-1}}{(2\pi)^{d-1}}
   \int^{\infty}_{0}dk k^{d-2}\sum_{n}g(n),
\end{equation}
where,
\[g(n)=\omega_{a_1,n}+2\omega_{a_2,n}-\omega_{b_1,n}-2\omega_{b_2,n}.\]
Now we are allowed to use the Abel-Plana summation formula, since
 we now expect the summand to satisfy the strict conditions \cite{Henrici} for the validity of this formula.
 That is, we expect any reasonable renormalization program for calculating any measurable physical quantity to
yield finite results. The Abel-Plana summation formula gives
\begin{eqnarray}\label{e8:main}
   E^{(0)}_{a}-E^{(0)}_{b}=\frac{L^{d-1}}{2}\frac{\Omega_{d-1}}{(2\pi)^{d-1}}
   \int^{\infty}_{0}dk k^{d-2}
  \hspace{2cm}\nonumber\\\times\left[\frac{-g(0)}{2}+\int^{\infty}_{0}g(x)dx+i
   \int^{\infty}_{0}\frac{g(it)-g(-it)}{\emph{e}^{2\pi
   t}-1}dt\right],
\end{eqnarray}
where $g(0)$  vanishes in this case due to our box renormalization.
The second term in the bracket, using suitable changes of variables,
becomes
\begin{widetext}
\begin{eqnarray}\label{e9:sec.term}
   && \dd\frac{a}
   {\pi}\int^{\infty}_0d\kappa\left(m_0^2+k^2+\kappa^2\right)^{1/2}+2\frac{L-a}{2\pi}
   \int^{\infty}_0d\kappa\left(m_0^2+k^2+\kappa^2\right)^{1/2}\nonumber\\
 &&\quad\dd-\frac{b}{\pi}\int^{\infty}_0d\kappa
   \left(m_0^2+k^2+\kappa^2\right)^{1/2}-2\frac{L-b}{2\pi}\int^{\infty}_0d\kappa
   \left(m_0^2+k^2+\kappa^2\right)^{1/2}=0,
\end{eqnarray}
where $\kappa$ for example in the first term denotes $\dd\frac{
n\pi}{a}$, obviously treated as a continuous variable. The above
calculation shows that this term is exactly zero. Therefore, only
the branch-cut term (the last term in Eq.~(\ref{e8:main})) gives
nonzero contribution and the final result is
\begin{eqnarray}\label{e10}
     E^{(0)}_{a}-E^{(0)}_{b}&=&-\frac{2L^{d-1}m_0^{^{(d+1)/2}}}{(4\pi)^{(d+1)/2}}
    \dd \sum^{\infty}_{j=1}\frac{1}{j^{(d+1)/2}}\nonumber\\
   && \times\left\{\frac{K_{_{(d+1)/2}}(2ajm_0)}{a^{(d-1)/2}}-\frac{K_{_{(d+1)/2}}
   (2bjm_0)}{b^{(d-1)/2}}+\frac{2K_{_{(d+1)/2}}[(L-a)jm_0]}{(\frac{L-a}{2})^{^{(d-1)/2}}}
   -\frac{2K_{_{(d+1)/2}}[(L-b)jm_0]}{(\frac{L-b}{2})^{^{(d-1)/2}}}\right\}
\end{eqnarray}
\end{widetext}
where $K_n(x)$ denotes the modified Bessel function of order $n$.
Using Eq.~(\ref{e2:E.Cas.}) for the Casimir energy and noting that
$K_n(x)$ is strongly  damped as $x$ goes to infinity, only the first
term remains when the limits are taken, and the result is
\begin{equation}\label{e11:res.z.term}
   E^{(0)}_{_{\mbox{\tiny Cas.}}}=-\frac{2L^{d-1}}{(4\pi)^{(d+1)/2}}
   \frac{m_0^{(d+1)/2}}{a^{(d-1)/2}} \sum^{\infty}_{j=1}
   \frac{K_{_{(d+1)/2}}(2ajm_0)}{j^{(d+1)/2}}.
\end{equation}
  If we set $d=3$, we have
\begin{equation}\label{e12}
   E^{(0)}_{_{\mbox{\tiny Cas.}}}=-\frac{L^2m_0^2}{8\pi^2a}\sum^{\infty}_{j=1}
   \frac{K_2(2ajm_0)}{j^2},
\end{equation}
with the following limits,
\begin{equation}\label{e13}
   E^{(0)}_{_{\mbox{\tiny Cas.}}}\rightarrow\left\{
   \begin{array}{ll}
        \dd\frac{-L^2}{8\pi^2a}\sum_{j}\frac{1}{2a^2j^4}
   =\dd\frac{-L^2\pi^2}{1440a^3} & \quad\mbox{as}\quad m_0\rightarrow0 \\
         \raisebox{-5mm}{$\dd\frac{-L^2}{8\sqrt{2}}(\frac{m_0}{\pi a})^{3/2}e^{-2am_0}$} &
         \raisebox{-5mm}{$\quad\mbox{as} \quad am_0\gg1.$}
   \end{array}\right.
\end{equation}
The results are in agreement with what exists in literature (see for
instance \cite{Svaiter,Mostepanenko}). It is interesting to note
that for the massless case, the result is, not surprisingly, exactly
half of the corresponding expression for the electromagnetic case.

For the $d=1$ case, Eq.~(\ref{e11:res.z.term}) becomes
\begin{equation}\label{e14}
   E^{(0)}_{_{\mbox{\tiny Cas.}}}=\dd-\frac{m_0}{2\pi}
    \dd \sum^{\infty}_{j=1}
  \dd\frac{K_1(2ajm_0)}{j},
\end{equation}
and its limits are
\begin{equation}\label{e15}
   E^{(0)}_{_{\mbox{\tiny Cas.}}}\rightarrow\left\{
   \begin{array}{ll}
         \dd-\frac{\pi}{24a} & \mbox{as}\quad m_0\rightarrow0 \\
         \raisebox{-5mm}{$\dd-\frac{1}{4}\sqrt{\frac{m_0}{\pi a}}e^{-2am_0}$} &
         \raisebox{-5mm}{$\mbox{as} \quad am_0\gg1,$}
   \end{array}\right.
\end{equation}
It is interesting to note that if we solve the massless case exactly
the branch-cut terms simplify to give
\begin{equation*}\dd
E^{(0)}_{_{\mbox{\tiny
Cas.}}}=-\frac{\pi}{a}\int_0^\infty\frac{t}{e^{2\pi
t}-1}dt=-\frac{1}{4a\pi}\zeta(2)=-\frac{\pi}{24a}
\end{equation*}
which is identical with our result for the massless limit, and is
reported for example, in Refs. \cite{Milton-p}.

\section{First-Order Radiative Correction}\label{sec3}

Now we calculate the next to the leading order (two loop quantum
correction) shift of the Casimir energy for a scalar field in
$\phi^{4}$ theory using the renormalized perturbation theory in
$1+1$ dimensions. As mentioned before, the main idea of our work is
that when a systematic treatment of the renormalization program is
done, the counterterms needed to retain the renormalization
conditions, automatically turn out to be position dependent. This,
as we shall see, will have profound consequences. However, our main
scheme of canceling the divergences  using counterterms and a few
input experimental parameters, is in complete  conformity with the
standard renormalization approach. To set the stage for the
calculations, we shall very briefly state the renormalization
procedure and conditions.

\subsection{Renormalization Conditions}
The $\phi^{4}$ Lagrangian Eq.(\ref{e1}), after rescaling the field
$\varphi=Z^{1/2}\varphi_{r}$, where $Z$ is called the field strength
renormalization, and the standard procedure for setting up the
renormalized perturbation theory, becomes (see for example
\cite{Peskin}),
\begin{eqnarray}\label{e17}
   &&{\cal L}(x)=\frac{1}{2}[\partial_{\mu}\varphi_{r}(x)]^{2}-\frac{1}{2}m^{2}
   \varphi_{r}(x)^{2}-\frac{\lambda}{4!}\varphi_{r}(x)^{4}\qquad\nonumber\\&&\hspace{1cm}+
   \frac{1}{2}\delta_{Z}[\partial_{\mu}\varphi_{r}(x)]^{2}
   -\dd\frac{1}{2}\delta_{m}\varphi_{r}(x)^{2}-\dd\frac{\delta_{\lambda}}{4!}\varphi_{r}(x)^{4},\nonumber\\
\end{eqnarray}
where $\delta_{m},\delta_{\lambda},\delta_{Z}$ are the counterterms,
and $m$ and $\lambda$ are the physical mass and physical coupling
constant, respectively. In this problem we are to impose boundary
conditions on the field at the walls. An alternative approach would
be to add appropriate external potentials to the Lagrangian so as to
maintain the boundary conditions on the fields. We will use the
first approach. Obviously the presence of nontrivial boundary
conditions breaks the translational invariance and hence momenta
will no longer be good quantum numbers. Therefore we find it easier
to impose the renormalization conditions in the configuration space.
For example, the standard expression for the two-point function is,
\begin{eqnarray}\label{2pfun.}
    &&\hspace{-1cm}\langle\Omega
    |T\{\phi(x_1)\phi(x_2)\}|\Omega\rangle\nonumber\\&=&\lim_{T\to\infty(1-i\epsilon)}
    \frac{\langle0|\int{\cal D}\phi\phi(x_1)\phi(x_2)e^{i\int_{-T}^T
    {\cal L} d^4x }\}|0\rangle}{\langle0|\int{\cal D}\phi e^{i\int_{-T}^T {\cal L}
    d^4x}|0\rangle}.
\end{eqnarray}
Since the birth of quantum field theory, as far as we know, the
assertion has always been that the above expressions can be expanded
systematically when the problem is amenable to perturbation theory.
For example, in the context of renomalized perturbation theory, as
indicated in Eq.(\ref{e17}), we can symbolically represent the first
few terms of the perturbation expansion of Eq.(\ref{2pfun.}) by
\begin{equation}\label{e18:renor.con1.}
   \raisebox{-4.5mm}{\includegraphics{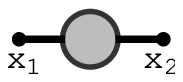}}=\raisebox{-3.3mm}{\includegraphics{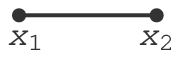}}
   +\raisebox{-3.7mm}{\includegraphics{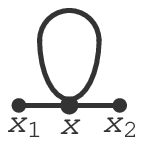}}+\raisebox{-3.9mm}{\includegraphics{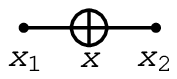}}+\dots.
\end{equation}
where \raisebox{-3mm}{\includegraphics{16}} refers to the
appropriate counterterm. It is obvious that the above expression
represents a systematic perturbation expansion, and most
importantly, all of the propagators on the right hand side should be
the one appropriate to the problem under consideration, that is they
should have the same overall functional form as the first term. Our
first renormalization condition is that the renormalized mass $m$
should be the pole of the propagator represented by the first term
in (\ref{e18:renor.con1.}). This implies the second and third
diagrams should cancel each other out in the lowest order, and this
in turn implies the cancelation of the UV divergences in that order,
and that the counterterms will in general turn out to be position
dependent. The renormalized mass $m$ will naturally turn out to be
position dependent as well. However, we only need to fix the value
of $m(x)$ at one position between the plates by our renormalization
condition. The exact functional dependence of $m(x)$ will then be
completely determined by the theory. That is, we insist the overall
structure of the renormalization conditions such as above, and the
counterterms appearing in them should be determined solely from
within the theory, and not for example be imported from the free
case. The equations are self deterministic and there is no need to
take such actions. Obviously we still need a few experimental input
parameters for the complete renormalization program, such as $m(x)$
for some $x$. Analogous expression and reasonings could be easily
stated for the four-point function.

To one-loop order the renormalization conditions derived from
Eq.~(\ref{e18:renor.con1.}) and its four-point counterpart, are
\begin{eqnarray}\label{e20:counterterms}
   &&\delta_{Z}(x)=0,\quad\delta_{m}(x)=\frac{-i}{2}\raisebox{-2.4mm}{\includegraphics{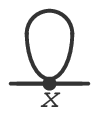}}
   =\frac{-\lambda}{2}G(x,x); \nonumber\\&&\mbox{and}
   \quad\delta_{\lambda}(x)=0,
\end{eqnarray}
respectively. Here $G(x,x')$ is the propagator of the real scalar
field and $x=(t,z)$. Obviously the counterterms automatically
incorporate the boundary conditions and are position dependent, due
to the dependence of the two and four-point functions on such
quantities. Now, the higher order contributions to the vacuum energy in the interval a1
(i.e. $z \in [\frac{-a}{2},\frac{a}{2}]$) is
\begin{eqnarray}\label{e21:vacc-pol}
   \Delta E_{a_1}= E^{(1)}_{a_1}+ E^{(2)}_{a_1}+\dots= \int^{a/2}_{-a/2}
   dz\langle\Omega|{\cal H}_{_I}|\Omega\rangle\qquad\nonumber\\=i\int^{a/2}_{-a/2} dz\left(\frac{1}{8}
   \raisebox{-7mm}{\includegraphics{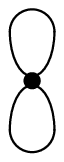}}+\frac{1}{2}
   \raisebox{-1mm}{\includegraphics{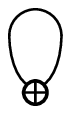}}+\frac{1}{8}\raisebox{-7mm}{\includegraphics{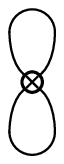}}+\dots
   \right),
\end{eqnarray}
where\raisebox{-3mm}{\includegraphics{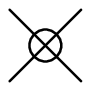}}$=-i\delta_{\lambda}(x)$
and
\raisebox{-1mm}{\includegraphics{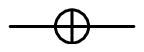}}$=i[p^{2}\delta_{Z}(x)-\delta_{m}(x)]$
refer to the counterterms. Accordingly, the ${\cal O}(\lambda)$
contribution to the vacuum energy is
\begin{eqnarray}\label{e22:vacc-ene}
   E^{(1)}_{a_1}&=&i\int^{a/2}_{-a/2} dz\left(\frac{1}{8}
   \raisebox{-7mm}{\includegraphics{8}} +\frac{1}{2}
   \raisebox{-1mm}{\includegraphics{9}}  \right)\nonumber\\&=&i\int^{a/2}_{-a/2}
   dz\bigg[\frac{-i\lambda}{8}G^{2}_{a_1}(x,x)\nonumber\\&&\hspace{10mm}
   -\frac{i}{2}  \delta_{m}(x)G_{a_1}(x,x)\bigg],\nonumber\\
\end{eqnarray}
where $G_{a_1}(x,x')$ is the propagator of the real scalar field in
region a1. Using Eqs.(\ref{e20:counterterms}) and
(\ref{e22:vacc-ene}), we obtain
\begin{eqnarray}\label{e23:E.Corr.}
   \hspace{-.6cm}E^{(1)}_{a_1}
   =\dd\frac{-\lambda}{8}\int^{a/2}_{-a/2}G^{2}_{a_1}(x,x)dz.
\end{eqnarray}

\subsection{ The Massive Case}\label{massive case}
As mentioned before, here we choose the Dirichlet boundary condition
on the plates. Then, after the usual wick rotation, the the
expression for the Green's function in the two dimensional Euclidean
space becomes
\begin{eqnarray}\label{e24:G.Func.}
   &&\hspace{-5mm}\dd G_{a_1}(x,x')=\nonumber\\&&\hspace{-5mm}\dd\frac{2}a\int\frac{d\omega}{2\pi}e^{\omega(t'-t)}\sum_{n}
   \frac{\sin\left[k_{a_1,n}(z+\frac{a}{2})\right]\sin
   \left[k_{a_1,n}(z'+\frac{a}{2})\right]}{\omega^{2}+k_{a_1,n}^{2}
   +m^{2}}.\nonumber\\
\end{eqnarray}
Using Eq.~(\ref{e24:G.Func.}) and Eq.~(\ref{e23:E.Corr.}) and
Carrying out the integration first over the space and then over
$\omega$, one obtains
\begin{eqnarray}\label{e25:E.bet.plats}
   &&\hspace{-.2cm}E^{(1)}_{a_1}=\frac{-\lambda}{8}\left[\frac{\pi^{2}}{a}\left(\sum_{n}
   \omega_{_{a_1,n}}^{-1}\right)^{2}
   +\frac{\pi^{2}}{2a}\sum_{n}\omega_{_{a_1,n}}^{-2}\right]\nonumber\\
   &&\hspace{0cm} =\frac{-\lambda\pi^{2}}{8a}\left[\left(\sum_{n=1}^\infty
   \frac{1}{\sqrt{\frac{n^2\pi^2}{a^2}+m^2}}\right)^{2}
   +\frac{am\coth am-1}{4m^2}\right].\nonumber\\
\end{eqnarray}
 According to  Eq.(\ref{e2:E.Cas.}) and Eq.(\ref{e3}) we have for
 the NLO correction
\begin{eqnarray}\label{e26:box-reg.}
    E^{(1)}_{\mbox{\tiny \mbox{\tiny Cas.}}}
    &=&\lim_{b/a\rightarrow\infty}\left[\lim_{L/b\rightarrow\infty}
   \left(E^{(1)}_{a}-E^{(1)}_{b}\right)\right],\nonumber\\
   \quad \mbox{where} \quad E^{(1)}_{a}&=&E^{(1)}_{a_1}+2E^{(1)}_{a_2},\quad
   E^{(1)}_{b}=E^{(1)}_{b_1}+2E^{(1)}_{b_2}.\nonumber\\
\end{eqnarray}
This computation is obviously complicated and plagued with a
multitude of infinities. As explained before using the usual
renormalization programs in conjunction with our box renormalization
scheme, should eliminate all of the infinities, as might be apparent
from the above equation. However, proper regularization schemes
should still be implemented and proper care taken when handling
these infinite expressions. For example, the summation appearing in
the squared form in the first term of the last part of
Eq.(\ref{e25:E.bet.plats}) is infinite. We want to use the
Abel-Plana formula to convert this sum into an integral. However
this sum does not satisfy the stringent requirements stated in the
Abel-Plana theorem  for such a conversion \cite{Henrici,Saharian}.
However our box renormalization scheme provides a solution: We first
expand the square as a double sum. Then we subtract these double
sums as indicated in Eq.~(\ref{e26:box-reg.}). Now we can expect
this new summand to satisfy the requirements for the Abel-Plana
theorem. Then all the infinities actually cancel and the result for
the two-loop correction reduces to (see Appendix for details):
\begin{widetext}
\begin{eqnarray}\label{e27:corr.cas.E}
     E^{(1)}_{a}-E^{(1)}_{b}=\nonumber\\
   &&\hspace{-2cm}\frac{-\lambda\pi^{2}}{8}
    \Bigg[f(a)-f(b)+2f(\frac{L-a}{2})-2f(\frac{L-b}{2})+\frac{2}{\pi}\left(B(a)-B(b)
   +2B(\frac{L-a}{2})-2B(\frac{L-b}{2})\right)
   \int_{0}^{\infty}\frac{ds}{\sqrt{1+s^{2}}}\Bigg],\nonumber\\
\end{eqnarray}
\end{widetext}
where,
\begin{eqnarray}  \label{A7}
f(a)=B(a)\left(\frac{B(a)}{a}-\frac{1}{am}\right)+\frac{\coth
(am)}{4m},
\end{eqnarray}
and $B(a)$, defined by the following expression
\begin{eqnarray}\label{eA4:diff.B}
   &&\hspace{-.8cm} B(a)=\dd
   2\int_{\frac{ma}{\pi}}^{\infty}\frac{1}{e^{2\pi t}-1}\frac{dt}
   {\sqrt{\frac{t^{2}\pi^{2}}{a^{2}}-m^{2}}},
\end{eqnarray}
refers to the so called branch-cut term in the Abel-Plana summation
formula and is a finite quantity. Note that the last integral in
Eq.~(\ref{e27:corr.cas.E}) seems to diverge so it must be properly
regularized, and this crucially depends on our box renormalization
program, as we shall explain below. However, before we engage in
this calculation, we want to raise an important point: If we were to
use the free counterterm in Eq.~(\ref{e22:vacc-ene}), as is
routinely done, this term would be absent, in addition to some minor
differences. Therefore, one would easily obtain finite results which
we like to dispute. Now let us proceed with our calculations. We
prefer to use a regularization scheme for this integral term which
is analogous to the zeta function regularization for the sums. That
is, in that expression we set the power of the integrand to
$-\frac{1}{2}+\alpha$ for the first two terms and
$-\frac{1}{2}+\alpha'$ for the remaining terms. In the final stage
we let $\alpha$ and $\alpha'$
 approach zero. Hence we will have
\begin{eqnarray*}
  &&2\frac{B(a)-B(b)}{\pi}\int_{0}^{\infty}(1+s^{2})^{-\frac{1}{2}
  +\alpha}ds\hspace{2cm}\\&&\hspace{1cm}
  +4\frac{B(\frac{L-a}{2})-B(\frac{L-b}{2})}{\pi}\int_{0}^{\infty}(1+s^{2})^{-\frac{1}{2}+\alpha'}ds\\&&
  =\frac{B(a)-B(b)}{\sqrt{\pi}\Gamma(\frac{-1}{2}+\alpha)}\Gamma(-\alpha)
  +2\frac{B(\frac{L-a}{2})-B(\frac{L-b}{2})}{\sqrt{\pi}\Gamma(\frac{-1}{2}+\alpha')}\Gamma(-\alpha').
\end{eqnarray*}
For $\alpha$ and $\alpha'$ sufficiently small, this expression
becomes
\begin{eqnarray}\label{e28}
   &&\hspace{-1cm}2\frac{B(a)-B(b)}{\pi}\left(\frac{-1}{2\alpha}+\ln2\right)\nonumber\\&&\hspace{1cm}
   +4\frac{B(\frac{L-a}{2})-B(\frac{L-b}{2})}{\pi}\left(\frac{-1}{2\alpha'}+\ln2\right).
\end{eqnarray}
Now, if
$\dd\frac{\alpha'}{\alpha}=-2\frac{B(\frac{L-a}{2})-B(\frac{L-b}{2})}{B(a)-B(b)}$
the infinities cancel. The cancelation of these divergent quantities
without any residual finite terms is the usual prescription in
regulated theories and this is what we have used \footnote{One may
argue that ambiguities always exist in problems where one has to
subtract infinite quantities, and the Casimir problems certainly
fall into this category. Two methods are in common use: First is the
analytic continuation techniques which, although usually yield
correct results, do not have a very solid physical justification and
also sometimes yield infinite results. Second is the regularization
schemes, which is what we have used. In the latter category when the
problem is regularized, one can make a systematic expansion of the
quantities in question in terms of the regulators. Then the terms
which tend to infinity when the regulators are removed and the
finite terms naturally appear separately. See for example Eq.
(\ref{e28}). What is almost invariably done is to adjust the
regulators so that the singular terms exactly cancel each other,
i.e. without extracting any extra finite piece from the difference
between the infinite quantities (see for example
\cite{Mostepanenko}). This is also apparent in the leading term for
the Casimir energy in Eq. (\ref{e9:sec.term}) where, as explained in
the Appendix, The four changes of variables are equivalent to
choosing four different cutoffs. One could have adjusted them so
that as usual the infinities cancel, but any finite term would
remain. However, the well known answer is obtained only when there
is no remaining extra finite term in this subtraction scheme. This
is the prescription that we have used. However, we do believe that
this is a subject that needs further study.}. Therefore the term in
question becomes,
\begin{equation}\label{e29}
   \frac{2}{\pi}\ln2
   \left(B(a)-B(b)+2B(\frac{L-a}{2})-2B(\frac{L-b}{2})\right).
\end{equation}
This result is obviously finite, and we believe it could not have
been obtained with any regularization or analytic continuation
schemes in common use, other than our box renormalization program.
Thus, Eq.~(\ref{e27:corr.cas.E}) becomes
\begin{eqnarray}\label{e31}
     &&\hspace{-1.2cm}E^{(1)}_{a}-E^{(1)}_{b}=\nonumber\\
   &&\hspace{-1.2cm}\frac{-\lambda\pi^{2}}{8}
    \left[f'(a)-f'(b)+2f'(\frac{L-a}{2})-2f'(\frac{L-b}{2})\right],
\end{eqnarray}
where $\dd f'(a)=f(a)+\frac{2\ln2}{\pi}B(a)$. This is the two-loop
radiative correction for the work done on the plates (or two points
in this case) while moving them from $(\frac{-b}{2},\frac{b}{2})$ to
$(\frac{-a}{2},\frac{a}{2})$. Now, in order to compute the Casimir
energy, proper limits must be taken, as indicated in
Eq.~(\ref{e26:box-reg.}).

Two particular limits are interesting to calculate. First is the
large mass limit. To calculate this limit it is convenient to make
the following expansion in the expression for $B(a)$,
Eq.~(\ref{eA4:diff.B}),
\begin{equation}\label{e34}
    \frac{1}{e^{2\pi t}-1}=\sum_{j=1}^{\infty}e^{-2\pi tj},
\end{equation}
then integration yields
\begin{equation}\label{e35}
    B(a)=\frac{2a}{\pi}\sum_{j=1}^{\infty}K_{0}(2amj)\quad{\buildrel {am\gg1 } \over
 \longrightarrow }\quad
    \sqrt{\frac{\pi}{2}}\frac{e^{-2amj}}{\sqrt{2amj}}.
\end{equation}
Using Eq.~(\ref{e35}) and Eq.~(\ref{e31}), Eq.~(\ref{e26:box-reg.})
gives
\begin{equation}\label{e41}
\begin{array}{ll}
   E^{(1)}_{_{\mbox{\tiny Cas.}}}& {\buildrel {am\gg1 } \over
 \longrightarrow }\dd\frac{-\lambda\ln2}{4}\lim_{\frac{b}{a}\rightarrow\infty}
 \Bigg\{\dd \lim_{\frac{L}{b}\rightarrow\infty}
   \Bigg[\dd \sqrt{\frac{a\pi}{m}}e^{-2am}-\sqrt{\frac{b\pi}{m}}e^{-2bm}\\
   & \dd +2\sqrt{\frac{(L-a)\pi}{2m}}e^{-(L-a)m}
   -2\sqrt{\frac{(L-b)\pi}{2m}}e^{-(L-b)m}\Bigg]\Bigg\}\\
   & \dd=\frac{-\lambda a\ln2}{4}\sqrt{\frac{\pi}{am}}e^{-2am}.
\end{array}
\end{equation}

In the  small mass limit it is easier to rewrite an expression for
$B(a)$, such that its integrand appears in dimensionless form,
\begin{equation}\label{ba1}
 B(a)=\dd
   \frac{2a}{\pi}\int_{1}^{\infty}\frac{1}{e^{2amt}-1}\frac{dt}
   {\sqrt{t^2-1}}.
\end{equation}
Then by expanding the integrand in Eq.~(\ref{ba1}) one finds,
$$B(a)\rightarrow\dd
\frac{1}{2m}-\frac{a}{\pi}\int_{1}^\infty\frac{dt}{\sqrt{t^2-1}}=\frac{1}{2m}-aS.$$
Note the explicit appearance of infrared divergences in this
equation which is a generic feature of these problems in 1+1
dimensions \cite{Coleman}. In this limit the first term in
Eq.~(\ref{e31}) due to region a1, for example, becomes
\begin{eqnarray}\label{e42}
    &&\hspace{-.5cm}a\left[\left(
    \frac{1}{2am}-S\right)\left[\left(\frac{1}{2am}-S\right)
    -\frac{1}{am}+\frac{2}{\pi}\ln2\right]+\frac{1}{4(am)^2}\right]\nonumber\\
    &&=a(S^2-\frac{2\ln2}{\pi}S)+\frac{\ln2}{\pi m}.
\end{eqnarray}
Hence, taking into account the analogous contributions from the
other regions, i.e. b1, a2 and b2, Eq.~(\ref{e26:box-reg.}) gives,
\begin{equation}\label{e42}
    E^{(1)}_{_{\mbox{\tiny Cas.}}}\rightarrow 0 \quad \mbox{as}\qquad
    m\rightarrow0.
\end{equation}
This shows that NLO for the Casimir energy in the massless limit is
zero. Most importantly the infrared divergences have also cancel
completely using our regularization program. This is in sharp
contrast to the analogous result that can be extracted from Refs.
\cite{Baron2, Caval} which is infinite.

\subsection{The Massless Case}
\begin{figure}[th]
\centerline{\includegraphics[width=7cm]{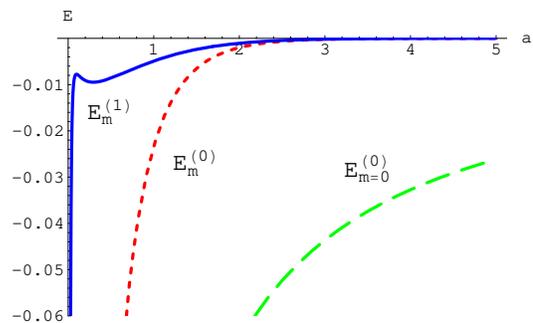}}
\caption{\label{fig:2} \small The Casimir energy and its first order
radiative correction for the $m=1$ and
 $m=0$ cases as a function of distance between the plates $a$.
Note that $E^{(1)}_{m=0}=0.$ We have used the following conventions
$\hbar=1$, $c=1$, and $\lambda=0.1 m^2$.}
\end{figure}

In the massless case it is sufficient to set the pole of the
propagator to zero, i.e. one can set $m=0$ in the
Eq.~(\ref{e24:G.Func.}). Now in the Eq.~(\ref{e23:E.Corr.}), after
space integration, carrying out the summation yields,
\begin{widetext}
\begin{eqnarray}\label{e37}
   E^{(1)}(a)&=&\frac{-\lambda}{8a}\Bigg[\dd\left(\int
   d\omega\frac{a\omega\coth(a\omega)-1}{2\omega^{2}}\right)^2
   +\frac{1}{2}\int
   d\omega' d\omega\frac{\omega'^{2}-\omega^{2}-a\omega'\omega^{2}\coth(a\omega')
   +a\omega'^{2}\omega\coth(a\omega)}{2\omega'^{2}\omega^{2}
   (\omega'^{2}-\omega^{2})}\Bigg]\nonumber\\
    &=&\dd
  \frac{-\lambda}{8}(P_{1}+aP_{2}),
\end{eqnarray}
\end{widetext}
where,
\begin{equation}\label{e38}
\begin{array}{ll}
   \dd P_{1}&\dd=\left(\int
   d p\frac{p\coth(p)-1}{2p^{2}}\right)^2,\\ P_{2}&\dd=\int
   \dd d p' d p\frac{p'^{2}-p^{2}-p'p^{2}\coth(p')
   +p'^{2}p\coth(p)}{\dd 4p'^{2}p^{2}(p'^{2}-p^{2})},\\
   & \dd p=a\omega\quad \mbox{and}\quad p'=a\omega'.
\end{array}
\end{equation}
Therefore,
\begin{equation}\label{e39:massless}
\begin{array}{ll}
   E^{(1)}_{_{\mbox{\tiny Cas.}}}
    \raisebox{0mm}{$=\dd\frac{-\lambda}{8}\left[a-b+
   2\frac{L-a}{2}-2\frac{L-b}{2}\right]P_{2}=0,$}
\end{array}
\end{equation}
since $P_{1}$ and $P_{2}$ are independent of $a,b$ and $L$. This
result is in exact agreement with the small mass limit calculated in
previous subsection. In figure~\ref{fig:2} we show our results for
the zero and first order Casimir energies for the massive and
massless cases.

Note that we are explicitly assuming $\delta_{m=0}(x) \neq 0$.
Although this does not happen to make any difference in 1+1
dimensions, using our prescription, we like to stress that this
quantity should not in general be a priori set to zero. This is in
contrast to the view expressed in for example
Refs.\cite{Symanzik,Krech,Ritberg, Geresd}. This is yet another
important counterexample for the validity of analytic continuation:
As is well known the massless limit of the analytic continuation of
the mass counterterm in $\phi^4$ theory is zero for space-time
dimension bigger than two. However one cannot renormalize the
massless theory without the mass counterterm (see for example
\cite{Peskin}).

\section{Conclusions}\label{sec4}
We have introduced a new concept in this paper. We have insisted
that the renormalization program should completely and
self-consistently take into account the boundary conditions or any
possible nontrivial background which break the translational
invariance of the system. We have shown that the problem is
self-contained and the above program is accomplishable. To be more
specific, there should be no need to import counterterms from the
free theory, or even supplementing them with the attachment of extra
surface counterterms, to remedy the divergences inherent in this
theory. In general this breaking of the translational invariance, is
reflected in the nontrivial position dependence of all the n-point
functions. As we have shown this has profound consequences. For
example in the case of renormalized perturbation theory, the
counterterms and hence the radiative corrections to parameters of
the theory, i.e. $m$ and $\lambda$, automatically turn out to be
position dependent. In this regard we disagree with authors who use
the former counterterms (see the Introduction for actual
references). Obviously we still need a few experimental input
parameters for the complete renormalization program, such as $m(x)$
for some $x$. However, the interesting point is that the theory then
completely determines $m(x)$.

Secondly we have used a supplementary renormalization scheme, which
is originally due to T.H. Boyer \cite{Boyer}, along side the usual
renormalization program. In computations of these sorts, there
usually appears infinities which can sometimes be removed by the
usual renormalization programs that often contain some sort of
analytic continuation. These procedures are sometimes ambiguous. Our
scheme is simply to confine the whole physical system in a box, and
to compute the difference between the values of the physical
quantity in question in two different configurations. Use of this
procedure removes all of the ambiguities associated with the
appearance of the infinities, and we use the usual prescription for
removing the infinities in the regulated theory. Moreover all of the
infrared divergences cancel each other. Using our method, we have
computed the zero and first order radiative correction to the
Casimir energy for the massive and massless real scalar field in 1+1
dimensions. For the zero order, our results are identical with what
exists in the literature. However, our first order results are
markedly different from those reported in Refs.\cite{Baron1,Caval}.
Our results for the massive case is different from theirs due to the
aforementioned conceptual differences. Moreover, we disagree with
their results for the massless case obtained as the limit of the
massive case, which is infinite in 1+1 dimensions. Our analogous
result is zero. The authors refer to the ``exact" results obtained
in Refs.\cite{Symanzik,Krech} as a verification of their massless
limits in 3+1. However, in the latter references the authors
effectively set $\delta_{m}$ equal to zero in their massless cases,
or its equivalent. This is our second main difference in approach to
the problem. As mentioned before, we believe that $\delta_{m}$
should not be arbitrarily set to zero even in the massless case,
since in that case the renormalization conditions can no longer be
fully implemented, although the theory is still in principle
renormalizable.

\begin{acknowledgments}
It is a great pleasure for us to acknowledge the useful comments of
Michael E. Peskin which helped us resolve one of the main
difficulties in our calculations. This research was supported by the
office of research of the Shahid Beheshti University. R.M. would
also like to thank the Qom University for financial support.
\end{acknowledgments}

\appendix*
\section{} \label{Calculation}

In this appendix we present the details of the calculations leading
to Eq.~(\ref{e27:corr.cas.E}). The Abel-Plana summation formula (see
for example \cite{Saharian}) is:

\begin{equation}\label{eA1:abel.plana}
    \sum_{n=1}^{\infty}f(n)=-\frac{f(0)}{2}+\int_{0}^{\infty}f(x)dx
    +i\int_{0}^{\infty}\frac{f(it)-f(-it)}{e^{2\pi t}-1}dt.
\end{equation}

In order to obtain first-order radiative correction, just as we
discussed in section [2] for leading term, we need to compute
$E^{(1)}_{a}-E^{(1)}_{b}$. Using  Eq.~(\ref{e25:E.bet.plats}), we
get
\begin{widetext}
\begin{eqnarray}\label{A2}
   &&\hspace{-.7cm}E^{(1)}_{a}-E^{(1)}_{b}=\frac{-\lambda\pi^{2}}{8}\dd\Bigg\{
   \dd\frac{am\coth am-1}{4am^2}-\frac{bm\coth
   bm-1}{4bm^2}+\frac{(L-a)\coth\frac{(L-a)m}{2}-1}{2(L-a)m^2}-\frac{(L-b)
   \coth\frac{(L-b)m}{2}-1}{2(L-b)m^2}\nonumber\\&&\hspace{.1cm}
   +\sum_{n,n'}\Bigg[\frac{1}{a}S(a,n)S(a,n')   -\frac{1}{b}S(b,n)S(b,n')+\frac{4}{L-a}S(\frac{L-a}{2},n)S(\frac{L-a}{2},n')
   -\frac{4}{L-b}S(\frac{L-b}{2},n)S(\frac{L-b}{2},n')\Bigg]\Bigg\},\nonumber\\
\end{eqnarray}
where $\dd S(a,n)=\left(m^2+\frac{n^2\pi^2}{a^2}\right)^{-1/2}$.
Using the Abel-Plana formula Eq.~(\ref{eA1:abel.plana}), and simple
changes of variables in the integrals, we obtain
\begin{eqnarray*}
   &&\hspace{-.7cm}E^{(1)}_{a}-E^{(1)}_{b}=\frac{-\lambda\pi^{2}}{8}\dd\Bigg\{\dd\frac{am\coth am-1}{4am^2}-\frac{bm\coth
   bm-1}{4bm^2}+\frac{(L-a)\coth\frac{(L-a)m}{2}-1}{2(L-a)m^2}-\frac{(L-b)\coth\frac{(L-b)m}{2}-1}{2(L-b)m^2}\\
   && \hspace{1.7cm}\dd+\sum_{n}\Bigg[-\frac{S(a,n)}{2am}
   +\frac{S(b,n)}{2bm}-\frac{2S(\frac{L-a}{2},n)}{(L-a)m}+\frac{2S(\frac{L-b}{2},n)}{(L-b)m}
   \\
   && \hspace{1.7cm}\dd+\frac{1}{\pi}\Bigg(S(a,n)-S(b,n)+2S(\frac{L-a}{2},n)-2S(\frac{L-a}{2},n)\Bigg)\int_{0}^{\infty}
   \frac{ds'}{\sqrt{m^{2}+s'^{2}}}\\
   && \hspace{+1.7cm}+\frac{B(a)S(a,n)}{a}-\frac{B(b)S(b,n)}{b}
   +\frac{4B(\frac{L-a}{2})S(\frac{L-a}{2},n)}{L-a}-\frac{4B(\frac{L-b}{2})S(\frac{L-b}{2},n)}{L-b}\Bigg]\Bigg\}.
\end{eqnarray*}
Using Eq.~(\ref{eA1:abel.plana}) again, and making appropriate
changes of variables to make the integrals dimensionless, all the
actual infinities cancel and we finally obtain
\begin{eqnarray}\label{A6}
     E^{(1)}_{a}-E^{(1)}_{b}=\nonumber\\
   &&\hspace{-2cm}\frac{-\lambda\pi^{2}}{8}
    \Bigg[f(a)-f(b)+2f(\frac{L-a}{2})-2f(\frac{L-b}{2})+\frac{2}{\pi}\left(B(a)-B(b)
   +2B(\frac{L-a}{2})-2B(\frac{L-b}{2})\right)
   \int_{0}^{\infty}\frac{ds}{\sqrt{1+s^{2}}}\Bigg].\nonumber\\
\end{eqnarray}
\end{widetext}
It is important to note that all these cancelations are easily
accomplished using our supplementary box renormalization scheme. On
a minor note, it is interesting to note that the changes of the
variables leading to the cancelation of infinities are,
surprisingly, equivalent to setting different cutoff regularizations
on the upper limits of the integrals. Equation (\ref{A6}) is our
main equation for the NLO Casimir energy, and appears in the text as
Eq.~(\ref{e27:corr.cas.E}), and is analyzed further there.


\begin{thebibliography}{9}
\bibitem{Casimir}
  H.B.G.~Casimir, Proc. Kon. Nederl. Akad. Wet. \textbf{51}  793 (1948).
\bibitem{Polder}
  H.B.G.~Casimir and D.~Polder, Phys. Rev. \textbf{73} 360  (1948).
  \bibitem{Svaiter}
    J. Ambj{\o}rn and S. Wolfram, Ann. Phys.  (N.Y.)  \textbf{147}  1 (1983).
    
    E. Elizalde, S.D. Odintsov, A. Romeo, A.A. Bytsenko, and S. Zerbini, 
    \emph{Zeta-regularization with Applications}, (World Scientific, 1994),

    N.J. Svaiter and B.F. Svaiter, J. Math. Phys.\textbf{ 32}  175 (1991),

    K.A. Milton, \emph{The Casimir effect}, (World Scientific, 2001),

    K.A. Milton, J.Phys. \textbf{A37} (2004) R209.
\bibitem{brw}
    M. Bordag, D. Robaschik, and E. Wieczorek, Ann. Phys. (N.Y.)  \textbf{165}   192 (1985).
\bibitem{bl}
    M. Bordag and J. Lindig, Phys. Rev. D  \textbf{58}  045003 (1998).
\bibitem{rsw}
    D. Robaschik, K. Scharnhorst, and  E. Wieczorek, Ann. Phys. (N.Y.)
    \textbf{174} 401  (1987).
\bibitem{bs}
    M. Bordag and K. Scharnhorst,  Phys. Rev. Lett. \textbf{ 81}  3815 (1998).
\bibitem{sx}
    S.-S. Xue., Commun. Theor. phys. (Wuhan) \textbf{11} 243  (1989).
\bibitem{zx}
    Tai-Yu Zheng and S.-S. Xue., Chin. Sci. Bull.  \textbf{38} 631 (1993).
\bibitem{rt}
    F. Ravndal, J.B. Thomassen, Phys. Rev. D  \textbf{63} 113007 (2001).
\bibitem{kr}
    X. Kong and F. Ravndal,  Phys. Rev. Lett.  \textbf{79}  545 (1997).
\bibitem{km}
    K. Melnikov, Phys. Rev. D \textbf{64} 045002 (2001).
\bibitem{ford}
    L.H. Ford, Proc. R. Soc. London A \textbf{ 368} 305 (1979).
\bibitem{kay}
    B.S. Kay,  Phys. Rev. D  \textbf{20} 3052  (1979).
\bibitem{toms}
    D.J. Toms,  Phys. Rev. D  \textbf{21} 2805 (1980).
\bibitem{lang}
    K. Langfeld, F. Schm\"{u}ser, and H. Reinhardt,  Phys. Rev. D \textbf{51} 765 (1995).
\bibitem{albu}
    L.C. de Albuquerque,  Phys. Rev. D \textbf{55} 7754 (1997).
\bibitem{Mostepanenko}
    M. Bordag, U. Mohideen, and V.M. Mostepanenko, Phys. Rep.  \textbf{353} 1 (2001).
\bibitem{Baron1}
    F.A. Baron, R.M. Cavalcanti, and C. Farina,   Nucl. Phys. B Proc. Suppl.
      \textbf{127} 118  (2004).
\bibitem{Baron2}
    F.A. Baron, R.M. Cavalcanti, and C. Farina, arXiv:hep-th/0312169 v1 (2003).
\bibitem{Graham}
    N. Graham, R. Jaffe, and Weigel, Int. J. Mod. Phys. A \textbf{17} 864
(2002).
\bibitem{Graham2}
    N. Graham, R. L. Jaffe, V. Khemani, M. Quandt, M. Scandurra, and H. Weigel, Nucl.Phys. \textbf{B645}
 49 (2002),

    N. Graham, R. L. Jaffe, V. Khemani, M. Quandt, M.
    Scandurra, and H. Weigel, Phys.Lett. \textbf{B572} 196 (2003),

    N.Graham, R.L.Jaffe, V.Khemani, M.Quandt, O. Schroeder, and H.Weigel, Nucl.Phys. \textbf{B677}
379 (2004).
\bibitem{Fosco}
    C.D. Fosco and N.F. Svaiter, J. Math. Phys. \textbf{42} 5185 (2001).
\bibitem{Nogu}
    J.A. Nogueria and P.L. Barbieri, Braz.  J. Phys.  \textbf{32} 798 (2002),

    M.I. Caicedo and N.F. Svaiter, J. Math. Phys. \textbf{45} 179 (2004),

    N.F. Svaiter, J. Math. Phys. \textbf{45} 4524 (2004),

     M. Aparico Alcalde, G. Flores Hidalgo, and N.F. Svaiter,
     J. Math. Phys. \textbf{47} 052303 (2006).
\bibitem{Caval}
     R.M. Cavalcanti and C. Farina,   arXiv:hep-th/0604200 (2006).
\bibitem{soliton}
     H.J. Vega, Nucl. Phys. \textbf{B115} 411 (1976),

     M.A. Lohe and D.M. O'Brien, Phys. Rev. D \textbf{23} 1771 (1981) ,

    N. Graham and R.L. Jaffe, Nucl. Phys. \textbf{B549} 516 (1999),

    E. Farhi, N. Graham, R.L. Jaffe, and H. Weigel, Nucl. Phys. \textbf{B585} 443 (2000),

    A.A. Izquierdo, W.G. Fuertes, G. Le\'{o}n, and J.M. Guilarte, Nucl. Phys.
     \textbf{B638} 378 (2002),

    A.A. Izquierdo, W.G. Fuertes, M.A. Gonz\^{a}lez Le\^{o}n, and J.M. Guilarte,
     Nuc. Phys. \textbf{B635} 525 (2002),


    A. Rebhan, P. van Nieuwenhuizen, and R. Wimmer,
     Nucl. Phys. \textbf{B648} 174 (2003),


    A.A. Izquierdo, W.G. Fuertes, M.A. Gonz\^{a}lez Le\^{o}n, and J.M.
    Guilarte, Nucl. Phys. \textbf{B681} 163 (2004).
    
     H. Nastase, M. Stephanov, P. van Nieuwenhuizen, and A. Rebhan, Nucl. Phys.
     \textbf{B542} 471 (1999),

\bibitem{rebhan}
    A. Rebhan and P. van Nieuwenhuizen, Nucl. Phys. \textbf{B508} 449 (1997),

   

    A.S. Goldhaber, A. Litvintsev, and P. van Nieuwenhuizen, Phys. Rev. D
     \textbf{64} 045013  (2001),

     
\bibitem{dashen}

R.F. Dashen, B. Hasslacher, and A. Neveu, Phys. Rev. D \textbf{10}
4114 and 4130 (1974); D \textbf{12} 2443 (1975).

\bibitem{moazzemi2}
  R. Moazzem, M. Namdar, and S.S. Gousheh, JHEP \textbf{09}(2007)029, arXiv:hep-th/0708.4127.
\bibitem{Bender}
    K.A. Milton, Ann. Phys. (N.Y.) \textbf{127} 49  (1980),

    S.K. Blau, M. Visser, and A. Wipf, Nucl. Phys. \textbf{B310} 163 (1988),

    C.M. Bender and K.A. Milton, Phys. Rev. D \textbf{50} 6547 (1994).

\bibitem{Boyer}
T.H. Boyer, Phys. Rev. \textbf{174} 1764 (1968).

\bibitem{Nest}
    V.V. Nesterenko, J. Phys. A: Math. Gen. \textbf{39} 6609 (2006).
\bibitem{Milton-p}
    V.V. Nestrenko and I.G. Pirozhenko, J. Math. Phys. \textbf{38} (12) 6265 (1997),

    K.A. Milton, Phys. Rev. D \textbf{68} 065020 (2003).
\bibitem{Peskin}
    Michael E. Peskin and Daniel V.Schroeder, \emph{An Introduction to Quantum
    Field Theory}, (Addison-Wesley, 1995)
\bibitem{Coleman}
S. Coleman, Phys. Rev. D \textbf{11} 2088 (1975).

\bibitem{Symanzik}
    K. Symanzik, Nucl. phys. \textbf{B190} 1 (1981).
\bibitem{Krech}
    M. Krech and S. Dietrich, Phys. Rev. A  \textbf{46}  1886 (1992).
\bibitem{Ritberg}
    T.V. Ritbergen and R.G. Stuart, Phys. Lett.  \textbf{B437} 201 (1998).
\bibitem{Geresd}
    G. von Gersdorff and A. Hebecker, Nucl. phys.  \textbf{B720} 211 (2005).
\bibitem{Henrici}
   P. Henrici, \emph{Applied and computational complex analysis}, Vol. 1, (Wiley, New
   York, 1984),

   E.T. Whittaker and G.N. Watson, \emph{A Course of Modern Analysis}, (Cambridge University
Press, 1958),
\bibitem{Saharian}
 A.A. Saharian, arXiv:hep-th/0002239; arXiv:hep-th/0708.1187.

\end{thebibliography}
\end{document}